\documentstyle[twocolumn,epsf,aps]{revtex}
\draft
\begin{document}

\title{Dynamical mean-field approximation for pair contact process
with a particle source}
\author{Attila Szolnoki$\dagger$}
\address{Research Institute for Technical Physics and Materials Science, 
H-1525 Budapest, POB 49, Hungary}
\address{
\centering{
\medskip \em
\begin{minipage}{15.4cm}
{}\qquad The one-dimensional pair contact process with a particle source is studied
by using dynamical cluster mean-field approximations with sites up to $n=12$. 
The results obtained for 
different levels of approximation become convergent especially for $n \ge 6$ 
and allow us to derive reliable extrapolations to the limit $n \to \infty$.
At the zero source limit, the critical point exhibits a discontinuity whose
magnitude vanishes with $1/n$.
The coherent anomaly analysis of data supports that the vanishing of order
parameter and density of isolated particles has the same critical behavior.
In contrast to an earlier prediction, the present approximation 
does not support the existence of critical behavior in the inactive phase
where the frozen density of isolated particles depends on the initial
state. 
\end{minipage}
}}
\maketitle

\narrowtext

The study of phase transitions from active fluctuating phases into absorbing 
states
has attracted considerable interest in the last decade \cite{ronbook,haye}. 
The static exponents of these phase transitions belong frequently to the 
directed percolation (DP) universality class, however, there 
are also examples for new universality classes \cite{dp}. 
Open questions are related to the 
necessary conditions that can destroy this robust universal behavior. 
One of the simplest model showing static DP behavior is the pair 
contact process (PCP) which was proposed to realize a system involving 
infinitely
many absorbing states \cite{iwan}. 
Recently, Dickman {\it et. al.} introduced
a modified PCP model to explore the robustness of DP transition \cite{acikk}. 
In this model, each site of the one-dimensional lattice is either vacant or
occupied by a single particle. A (randomly chosen) pair of nearest-neighbor
particles is annihilated with a probability $p$ or an additional particle
is created around the given pair with a probability $1-p$ if it is not
forbidden by double occupancy.
In the extended model, an external particle source is introduced that
attempts to insert {\it isolated} particles with a rate of $h$. 
This system exhibits an active phase when $p$ is smaller than a critical
value, $p_c$. For $p \ge p_c$ the system evolves into a frozen (absorbing)
state where the nearest-neighbor pairs are absent. 
Further details of the model can be found in Ref.~\cite{acikk}.
This extended model was studied by Monte Carlo (MC) simulations and
dynamical cluster mean-field approximation for quite large cluster
sizes (ranged from $n=2$ to $6$). Some disturbing behaviors, however,
remained unsolved. For example, the analytical predictions tend not
monotonously toward the MC results for $h>0$ when the cluster size
is increased. Furthermore the analytical results shows a discontinuity
in the variation of critical point $p_c$ if $h \to 0$. 
This observation is surprising because the present approximation has
proved to be satisfactory in many cases for $n \le 6$ \cite{satis}.
It is expected that the further increase of cluster size will
resolve these discrepancies.

In this Brief Report, we discuss the results of dynamical mean-field
approximations for cluster sizes as large as $n=12$. The present approach
allows us to give more accurate extrapolations to the limit
$n \to \infty$. Using this large number of data (from $n=2-12$) we
can also improve the prediction of coherent anomaly method (CAM)
\cite{cam} when considering the critical exponents 
around the transition point. Furthermore, we could study the frozen 
density of isolated particle concentration toward the system evolves
in the inactive phase \cite{port}.

To identify the transition point we have evaluated the $p$-dependence of
the density of pairs of particles $\rho_2$ (henceforth considered as
order parameter) as well as the density of isolated particles $\rho_1$.
The transition point is determined by the zero-point of $\rho_2$ for
$h=0$ and by the breaking point of the function $\rho_1(p)$ for $h>0$.
This latter criteria was particularly useful because of the extremely
low value of $\rho_2$ in a large interval of $p$.

We shall not describe the details of the mean-field technique because
it has already been applied and demonstrated for the PCP model by
several authors previously \cite{acikk,port,henkel}.
The derivation of the hierarchy of equations of motion for the
configuration probabilities on the $n$-site clusters becomes complicated
for large $n$. The technical difficulties increase drastically when 
enlarging the size of clusters. Traditionally, the set of equation 
of motion is solved numerically in the stationary states. Despite the
slow convergency toward the stationary solution for large $n$,
the numerical integration of the master equations seems to be a more
efficient method to find the stationary probability of configurations
than the traditional Newton-Raphson method.
Using the numerical integration method we could determine all the 
configuration probabilities
appearing on the 12-site cluster. (As an example, to get a data in
last row of Table~I requires 4-week running on a personal computer.)  
It is worth mentioning that very recently this method is used successfully
for the consideration of a stochastic sandpile model \cite{ronsand}. 

\begin{figure}
\vspace{3cm}
\centerline{\epsfxsize=8.0cm
            \epsfysize=8.0cm
                  \epsfbox{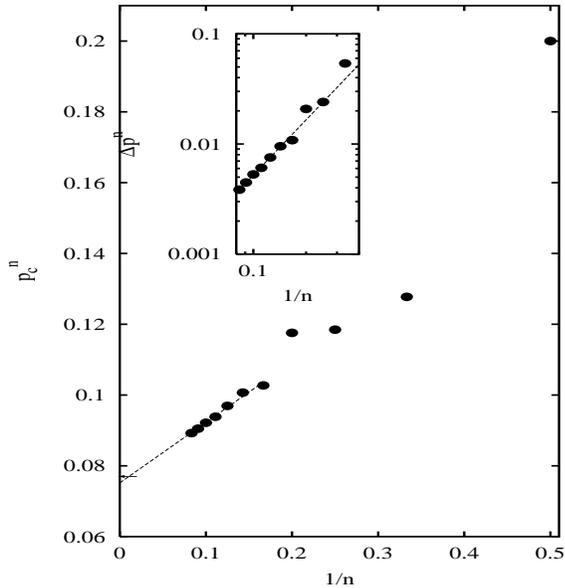}
                  \vspace*{2mm}  }
\caption{Predictions for $p_c$ at different levels of approximation
($n=2, \ldots, 12$) at $h=0$. Dashed line is the linear fit for 
$n=6, \ldots, 12$ data and arrow shows the MC data.
Insert: Discontinuity in $p_c$ at $h \to 0$ as a function of levels of 
approximation. $n=3, \ldots, 12$-point levels are plotted.
Dashed line corresponds to a power law with an exponent $\omega \approx 1.666$.}
\label{fig:pcjump}
\end{figure}

\begin{figure}
\vspace{3cm}
\centerline{\epsfxsize=7.0cm
            \epsfysize=7.0cm
                  \epsfbox{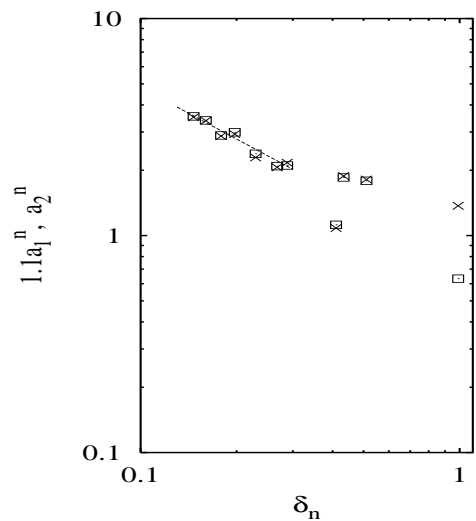}
                  \vspace*{2mm}  }
\caption{CAM scaling of the critical mean-field coefficients for the order 
parameter ($a_2$, open squares) and the 
density of isolated particles ($a_1$, crosses).
Dashed lines correspond to the $\beta = 0.276$ exponent.
The values of $a_1^n$ are multiplied by $1.1$.}
\label{fig:cam}
\end{figure}

\begin{figure}
\vspace{3cm}
\centerline{\epsfxsize=7.0cm
            \epsfysize=7.0cm
                  \epsfbox{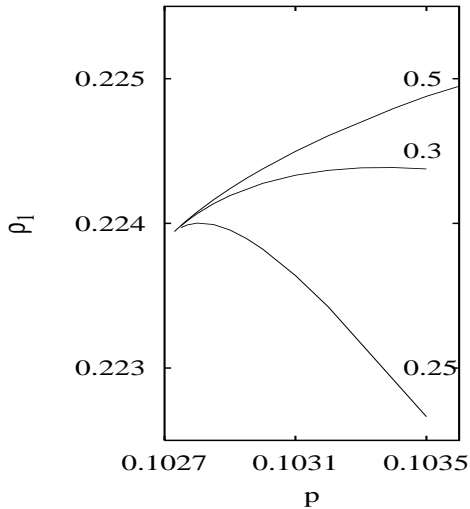}
                  \vspace*{2mm}  }
\caption{The stationary value of particle concentration in the inactive 
phase at $6$-point level at $h=0$. 
The initial concentrations for the different curves are indicated.}
\label{fig:inac}
\end{figure}

Our results are plotted in Figs.~\ref{fig:pcjump}-~\ref{fig:inac}. 
Table~I summarizes the predictions of all levels of
approximations for $p_c$ and for the density of 
isolated particles $\phi_{nat}$ at the critical point at different
values of $h$. 
Figure~\ref{fig:pcjump} demonstrates clearly that the predictions for
$p_c$ become monotonously convergent for $n \ge 6$. The linear fit 
for $n>5$ data gives $p_c = 0.075$ in close agreement with the MC
data ($p_c^{MC} = 0.077$). The same good convergence
can be obtained in the presence of external source. 
The comparisons of linear fit with MC data are also listed in Table~I 
(MC data are taken from Ref~\cite{acikk}).

Surprisingly, although the $n$-site approximations converge to MC data
at {\it arbitrary} value of $h$, a jump can be observed
in $p_c$ when considering the limit $h \to 0$ for all $n$. 
The magnitude of this jump is defined as
\begin{equation}
\Delta p^n = \lim_{h \to 0} p_c^{n}(h)  - p_c^{n}(h=0)  \,\,.
\label{eq:jump}
\end{equation}
The inset of Fig~\ref{fig:pcjump} shows $\Delta p^n$ for different levels 
of the approximation. This log-log plot suggests that the jump decreases
as a power low function, i.e., $\Delta p^n \propto n^{-\omega}$ with
an exponent $\omega \approx 1.666$. The explanation of this value of
$\omega$ remains to be clarified.  
This discontinuity of $\Delta p^n$ may also be observed for higher 
dimension versions of the model \cite{szol}. 

A distinct improvement of convergency can also be observed for higher values
of $n$ if we consider the density of isolated particles $\phi_{nat}$ at 
$p_c$. The extrapolations, based on predictions of $n>5$ data, 
are in excellent agreement with MC data.

Such a large number of approximations makes possible to apply the CAM 
analysis introduced by Suzuki \cite{cam}. 
In the vicinity of the critical point the order parameter ($\rho_2$) and
the density of isolated particles ($\rho_1$) are estimated by 
$\rho_2 \propto a_2^n (p_c^n-p)$ and
$\rho_1-\phi_{nat} \propto a_1^n (p_c^n-p)$, where $p_c^{n}$  
denotes the prediction for the critical point at $n$-point level.
To estimate the $\beta$
critical exponents we have plotted the amplitudes of mean-field results
($a_1^n$ and $a_2^n$) as a function of
\begin{equation}
\delta_n = (p_c^{n}/p_c)^{1/2} - (p_c/p_c^{n})^{1/2}  \,\,,
\label{eq:cam}
\end{equation}
where $p_c$ denotes the result of MC simulation.
Figure~\ref{fig:cam} shows the mean-field amplitudes in the absence
of external source ($h=0$) for different levels of approximation.

Using different subsets of CAM-data the estimated exponents are 
$0.221, 0.246, 0.276,$ and $0.312$. 
The dashed lines correspond to the $\beta = 0.276$ DP exponent.
Although the error bar is still large but the tendency to DP is 
straightforward. 
These results confirm
the earlier MC observations \cite{gez}. Namely, the order parameter and
the density of isolated particles can be described by the same power laws
when the system approaches the critical point from the active phase.
At the same time the CAM analysis of mean-field data helps 
to understand why the exponents agree. Figure~\ref{fig:cam} demonstrates
that these two quantities are proportional to each other in the vicinity
of critical point. (In this plot $1.1 a_1^n$ is compared with $a_2$).

In the presence of source ($h>0$) the earlier MC simulation suggested
a slightly modified $\beta = 0.287$ exponent \cite{acikk}. Our mean-field 
data become also convergent for $n \ge 6$
suggesting an exponent close to DP class ($\beta =0.274 \pm 0.038$ at $h=10$). 
However, the region in $\delta_n$
is so small that 
the error bar of the estimated exponent is four times larger than the 
predicted change of MC result. Therefore,  
this method is incapable to distinguish such close exponents.

As noted above, the large-cluster approximation makes possible to
test the prediction of {\it static} critical behavior of particle 
concentration in the 
{\it inactive} phase \cite{port}. Such a critical behavior of a static 
quantity
in the inactive phase has already been observed in a different model
\cite{lip}.
In a previous MC study of PCP model it was suggested that the density of
isolated particles follows a power law 
$\rho_1^{nat} - \rho_1 \propto (p-p_c)^{\beta_1}$ for $p > p_c$
in the absence of external source \cite{port}. 
Following Marques {\it et al.} \cite{port}, we have calculated 
the absorbing state selected by the system's dynamics starting from
the same homogeneous initial state. However, different initial particle 
concentrations were chosen to study the robustness of the final absorbing
state in the inactive phase. The initial concentration is ranged from 
$\phi_n$ to $0.99$. The results of 
$6$-point approximation are plotted in Fig.~\ref{fig:inac}
for different values of initial concentration at $h=0$. 
It suggests that the 
absorbing state 
selected by the system's dynamics depends on the initial condition.
Similar behavior may be observed for all levels of approximation.
The observed deviation of curves increases further by increasing $p$ 
and the level of approximation.
Obviously, the stationary value of concentration becomes 
independent of the initial condition in the active phase.

We have tried to apply CAM analysis on the data obtained from the 
{\it same} initial concentration for different levels of 
approximation. Even in this case, the irregularity
of mean-field coefficients at different levels of approximation
does not allow to extract critical exponent. Shortly, the 
present approximation does not support the existence of static 
critical behavior in the inactive phase.
In the light of this prediction further intensive MC simulations
are suggested to clarify 
the possible existence of {\it natural absorbing states}.

In summary, we have demonstrated that the present dynamical
mean-field approximations (for large $n$) yield adequate extrapolation 
to the limit $n \to \infty$ even in the presence of external particle
source that causes nonanalytical behavior in the limit $h \to 0$.
According to this approximation the discontinuity of $p_c$
displays power-law decay. 
The application of CAM analysis demonstrates that the densities of
nearest-neighbor pairs and isolated particles are proportional to each 
other in the vicinity of the critical point and 
supports the same critical exponents of these quantities. This behavior
is certainly related to the fact that the extinct pairs can leave
extra solitary particles behind.
Although our analysis is proved to be a useful tool for investigating
stationary states in the active phase it
does not confirm the occurrence of an
unambiguous critical behavior in the inactive phase where 
the composition of the frozen state depends on the initial state.
We hope that the obscurity of stationary inactive state will stimulate
further MC simulations and theoretical investigations. 

\vspace{0.5cm}

The author wishes to thank Ron Dickman, Gy\"orgy Szab\'o, and M. C. Marques 
for stimulating discussions. This research was supported by the Hungarian 
National Research 
Fund (OTKA) under Grant No. F-30449 and Bolyai(BO/0067/00).

\noindent$^\dagger$ Electronic address: szolnoki@mfa.kfki.hu

\centering{
\medskip \em
\begin{minipage}{15.4cm}
{}\qquad\begin{table}
\caption{\sf results of $n$-site approximations}
\begin{center}
\begin{tabular}{|c|l|l|l|l|l|l|} 
$n$ &$p_c (h\!=\!0)$&$p_c (h\!=\!0.1)$&$p_c (h\!=\!10)$&$\phi (h\!=\!0)$&$\phi (h\!=\!0.1)$&$\phi (h\!=\!10)$ \\
\hline\hline
2 &0.2    &0.66667&0.6667 &0      &0.5   &0.5    \\
3 &0.12774&0.1820 &0.182  &0.23   &0.4615&0.4615 \\
4 &0.11846&0.1502 &0.1573 &0.21665&0.5   &0.5    \\
5 &0.11757&0.1440 &0.1579 &0.17989&0.4580&0.477  \\
6 &0.10272&0.1166 &0.1353 &0.22393&0.4383&0.4482 \\
7 &0.10069&0.1148 &0.1323 &0.21840&0.4450&0.4511 \\
8 &0.09692&0.1094 &0.1265 &0.22121&0.4390&0.4476 \\
9 &0.09387&0.1044 &0.1224 &0.22879&0.4356&0.4442 \\
10&0.09218&0.1027 &0.1198 &0.22838&0.4351&0.4438 \\
11&0.09048&0.1004 &0.1175 &0.23120&0.4325&0.4418 \\
12&0.08925&0.0989 &0.1156 &0.23269&0.4316&0.4416 \\
\hline
$\infty$    &0.075      &0.080       &0.095  &0.242   &0.422   &0.433\\
\hline\hline
{\small SIM}&0.077091(5)&0.086272(15)&0.09785&0.241(1)&0.421(1)&0.433\\
\end{tabular}
\end{center}
\label{table:nsite}
\end{table}
\end{minipage}
}

\end{document}